\documentclass[12pt]{article}

\usepackage{authblk}

\usepackage{amsmath}
\usepackage{latexsym}
\usepackage{amsmath}
\usepackage{amssymb}
\usepackage{amsfonts}
\usepackage{bm} 
\usepackage{eurosym} 
\DeclareUnicodeCharacter{20AC}{\euro}

\usepackage{graphicx} 
\usepackage{graphics}
\usepackage{caption}
\usepackage{subcaption}
\usepackage{float}
\usepackage[section]{placeins}
\usepackage{pdfpages}
\usepackage{pdflscape}

\usepackage{booktabs} 
\usepackage{array} 
\usepackage{paralist} 
\usepackage{verbatim} 

\title{Electric or gasoline: a simple model to decide when buying a new vehicle}

\author[1]{Ren\'e Ledesma-Alonso \thanks{\texttt{rene.ledesma@ciencias.unam.mx}}}
\affil[1]{\small
Facultad de Ciencias\\
Universidad Nacional Aut\'onoma de M\'exico\\
Av. Universidad 3000, Circuito Exterior S/N Alcald\'ia Coyoac\'an, C.P. 04510 Ciudad Universitaria, Ciudad de M\'exico, M\'exico.}

 \author[2]{Guillermo Becerra-Nu\~nez \thanks{\texttt{catedra.guillermo@uqroo.edu.mx}}}
 \affil[2]{Department of Engineering,\\
CONAHCYT - Universidad Aut\'onoma del Estado de Quintana Roo,\\
Chetumal 77019, Quintana Roo, M\'exico.}

\date{}

\begin{document}
\maketitle

\begin{abstract}
In this work, a simple methodology to follow the behavior of motorized urban vehicles, from the point of view of personal finances, is presented.
Including the acquisition of a new vehicle, the analysis considers the driving distance per week, the average speed, the time spent at rest due to traffic conditions, the evolution of gasoline and electric energy prices, maintenance and services, and local taxes.
Herein, two low-range compact vehicles were chosen and compared: one powered by combustion of gasoline, and the other by electric energy stored in batteries.
Historical data and trend projections, according to inflation and prices evolution, are taken into consideration.
The developed model may help to select adequately a new vehicle, according to the user's needs.
A good choice depends strongly on the usage and traffic conditions, the electric vehicle being suitable for large weekly driving distances and heavy traffic, whereas the gasoline vehicle is preferred for short distances and light traffic.
The expenses of the vehicles are compared through time, with different scenarios envisaged according to the user's resolution to keep the vehicle for the entire lifespan or to sell it quickly.
\end{abstract}

\section{Introduction}

During the previous century, humanity's main development was based on the use of hydrocarbons, which has had a serious impact on the environment, in addition to the political issues associated with oil consumption.
Similarly, the transportation sector has been considered one of the main contributors to air pollution, accounting for 14$\%$ to 25$\%$ of the total greenhouse gas emissions~\cite{Newsham2018,Ilie2019,Lakshmi2023}.
Vehicles powered by internal combustion engines are particularly recognized for generating the highest emissions~\cite{Wang2023}.
Due to these problems, actual discussions are taking place regarding the potential ban on the sale of combustion engine vehicles that employ fossil fuels, starting from 2035 within the European Union~\cite{CNNmar2023}.

Different technologies have been developed to address reduce pollutant emissions and minimize fossil fuel consumption in combustion engines.
An alternative approach proposes the use of fuel mixtures in these engines, such as blending hydrogen with fossil fuels~\cite{Wang2023,Wang2022,Liu2023} and bio-fuels~\cite{Otchere2020, Singh2020, Becerra2021}.
In recent years, significant advances have been made on the development of various technologies for commercial vehicles.
One remarkable example is the emergence of hybrid electric vehicles (HEVs), which combine traditional internal combustion engines with electric motors, for propulsion.
The HEVs have proven to be highly efficient and have gained popularity, due to their ability to reduce fuel consumption and emissions~\cite{Hong2021, Tintu2023, Zheng2023}. 
The introduction of hybrid electric vehicles has marked a significant milestone in the past three decades, playing a crucial role in the continuous evolution and commercialization of electric vehicles.
However, despite the achieved progress, there are still noticeable challenges that need to be addressed, particularly in the areas of energy storage systems and charging infrastructure, as well as the way to produce electric energy, due to the type of used energy source and its possible pollutant residues.

Furthermore, considerable progress, over the past decade, has been achieved in the ambit of electric vehicles.
The range of energy autonomy has experienced a substantial boost, expanding from 100--150 $\textrm{km}$ to more than 400 $\textrm{km}$~\cite{Mierlo2021}.
Battery storage capacity has undergone a significant improvement, due to the development of different types of electrochemical cells, of increasing from 110 $\textrm{W}\cdot\textrm{h}/\textrm{kg}$ in 2010 to 250 $\textrm{W}\cdot\textrm{h}/\textrm{kg}$ in 2020~\cite{Ntombela2023}.
Similarly, throughout the same decade, there has been a parallel enhancement of energy density within the battery, rising from 310 $\textrm{W}\cdot\textrm{h}/\textrm{L}$ to more than 500 $\textrm{W}\cdot\textrm{h}/\textrm{L}$~\cite{Mierlo2021}, whereas the cost of batteries has experienced a remarkable reduction, dropping from $1000\, \textrm{USD}/\textrm{kW}\cdot\textrm{h}$ to $150\, \textrm{USD}/\textrm{kW}\cdot\textrm{h}$~\cite{BatteriesEERE}.
It is also worth highlighting that electric vehicles offer a substantial advantage in terms of energy conversion efficiency, when compared to combustion vehicles~\cite{Bairwa2023}. 
The electric motor of an electric vehicle converts approximately 75$\%$ of the energy it consumes into mechanic energy (without including any conversion efficiency from its power generation source), whereas a conventional combustion engine only achieves around 20$\%$ efficiency.
Even considering a low power generation efficiency of around 40$\%$, for a combined cycle plant, a superior energy conversion efficiency of electric vehicles (of around 30$\%$) is expected.
Thus, the direct consequences of an electric vehicle employment are the reduction of energy waste and that of its environmental impact.

Several previous studies have addressed the economic comparison between different types of vehicles.  
For instance, an economic comparison between gasoline and hybrid vehicles has been presented more than two decades ago~\cite{Lave2002}. 
The authors conclude that the gasoline vehicle remains cheaper, even including the social cost of emissions and pollutants. 
Nevertheless, that analysis is based on static prices of fuel (for the year 2002), which does not reflect the real economic evolution and trends. 
More recently, the competitiveness of electric vehicles against gasoline and hybrid vehicles has been tested again in the U.S. and Germany~\cite{Simeu2018}, for which electric vehicles have economic advantages over the others, when large driving distances are envisaged.
However, in this study, the prices of electricity and fuel, initial costs and ownership taxes have been extrapolated to a single year (2025), neglecting the dynamics of the economic system.

As well, an excel tool has been developed for the U.S., including historical data of the fuel and electricity prices from 2015 to 2020, as well as the fuel or electric energy consumption per distance and the monthly driving distance~\cite{Schmidt2021}.
Despite of the simplicity of their model and the energy cost calculator that the authors present, a systematic analysis of the effect of the parameters on the total costs, and a comparison between the gasoline and electric outcomes, has not been developed.
Finally, it is worthy to mention that other studies have been focused on the evaluation of greenhouse gas emissions of different vehicles, which is directly related with static costs and savings, or energy consumption.
In particular, driving cycles, urban, rural and combined region conditions, and electric energy generation sources have been considered~\cite{Yuksel2016}.
Also, the effect of driving cycles, stochastic speed profiles and different percentages of connected-autonomous vehicles on the energy consumption have been tested 
~\cite{Islam2020}.

Herein, we will develop a simple model evaluate the total cost, though time, when acquiring a new vehicle.
It has been applied to the economic data of Mexico City, since we have direct access to them, and a comparison between the less expensive gasoline and electric vehicles, that can be acquired in Mexico, is performed.
Nevertheless, the general approach of the model allows it to be easily adapted to any city in the world, just by modifying the specific trends and functions of the economic variables, and to any specific vehicle as well.

Section 2 includes a description of the variables involved and the way in which the energy consumption of gasoline and electric vehicles is modeled, as well as the conversion to prices.
Section 3 describes economic variables that are required to evaluate the current cost (depreciation), services and maintenance of the vehicle.
In Section 4, all variables and parameters are integrated into a pair of equations, that describe the total accumulated expenditure and capital recovery for a vehicle in a given time.
A brief description of the numeric calculations is given in Section 5.
In Section 6, the results for 2 low-cost vehicles, one gasoline-powered and the other electric, are presented and compared, for the conditions of Mexico City.
Finally, Section 7 presents the conclusions drawn for this study.

\section{Motion characteristics and vehicle capabilities}
\label{Sec:Motion}

Many factors, including the topography, road map, and weather conditions, affect the performance and energy consumption of a vehicle.
Herein, with the aim of establishing a simple method to quantify vehicle usage for comparison purposes, we have selected general variables, which can be measured easily and with basic tools.
For instance, using a satellite navigation software and a clock, one can measure the travel distance and the elapsed time of a journey, whereas employing a chronometer allows to estimate the duration of the total stop time (zero velocity of the vehicle).
With these measurements, one can determine the average speed of the vehicle, by dividing the distance by the difference between the journey time and the total stop time.
Other situations, such as the variations of road slope, will affect the behavior of the vehicle, which can be addressed by averaging the fuel and energy consumption.
In the following, we will explain how these simple variables are involved in the presented model.

Consider a vehicle used to travel a driving distance $\ell$, given in $\textrm{km}/\textrm{week}$, at an average speed during motion $U$, in $\textrm{km}/\textrm{h}$. 
The value of $U$ relies on the speed limits and the density of traffic lights, among other factors which include the driving habits and route selections.
The time spent in the vehicle $T$, in $\textrm{h}/\textrm{week}$, which depends not only on the driving distance $\ell$ and the average speed $U$, but also strongly on the density of vehicles and the occurrence of unexpected events, should be in the range $T_{\textrm{min}}\leq T<T_{\textrm{max}}$.
Likewise, the minimum time, $T_{\textrm{min}}$, is measured when no delays occur during all the trajectories in a week, and it is equal to $T_{\textrm{min}}=\ell/U$.
On the other hand, the maximum time is the duration of a week in hours, $T_{\textrm{max}}=7\cdot24=168\, \textrm{h}/\textrm{week}$.
It might be impossible to attain $T_{\textrm{max}}$, since the employment of the vehicle is normally disrupted, due to energy replenishment (chemical or electric) or driving breaks (human basic needs).
The range of a vehicle is taken as the maximum distance  $\ell_{range}$, that it can travel without any disruption, before  exhausting its stored energy.

In order to analyze the monthly evolution of energy consumption and the expenses, that are related to the acquisition and usage of the vehicle, we will define the time $t$, given in $\textrm{year}$, as a discrete variable described by:
\begin{eqnarray}
t&=y+\frac{m}{12} \ ,
\end{eqnarray}
where $y$ is the year and $m$ is the number of elapsed months since the beginning of the year $y$.
Under the same description, the time $t_0$, at which a vehicle is acquired, reads:
 \begin{eqnarray}
 t_0&=y_0+\frac{m_0}{12} \ ,
 \end{eqnarray}
where $y_0$ is the year of acquisition and $m_0$ are the elapsed months since the beginning of the year $y_0$.
Herein, we will consider $y_0$ as the reference year, which will be also employed for the calculation of different trends.
Additionally, the number of months $M$ and years $Y$ elapsed from $t_0$ to $t$ are thus given by:
\begin{subequations}
\begin{eqnarray}
M&=12\left(t-t_0\right) \ , \\
Y&=\lfloor t-t_0\rfloor \ .
\end{eqnarray}
\end{subequations}
Finally, the average number of weeks of a month is given by $w=365/(12\cdot 7)$. 

\subsection{Gasoline consumption and price}

\begin{figure}
\centering
\includegraphics[width=0.50\textwidth,trim=35mm 80mm 30mm 84mm,clip]{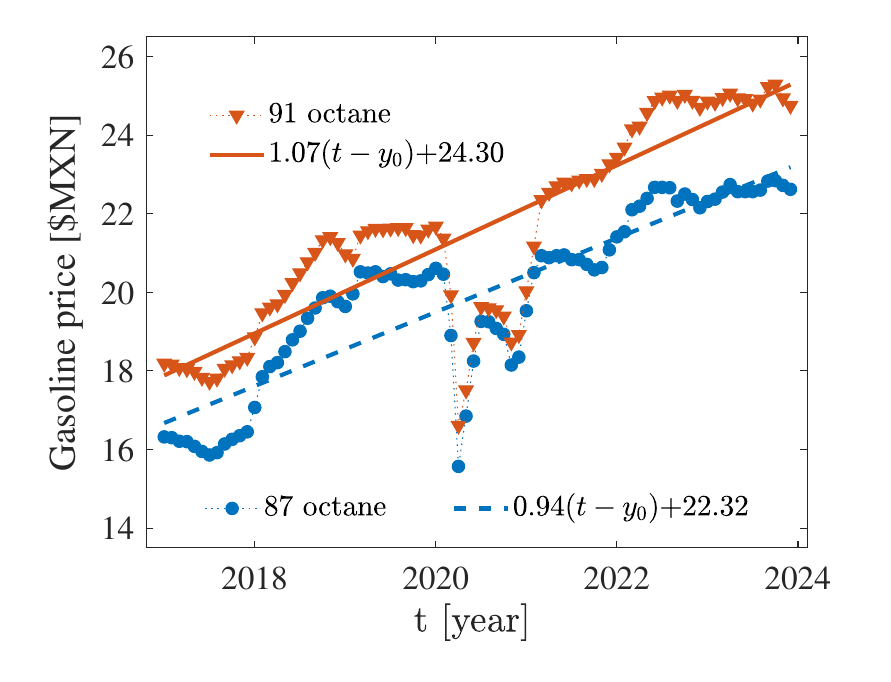}
\caption{Price of the two types of gasoline found in Mexico City. The reference year is $y_0=2023$, and linear trends are shown, with $a_{\textrm{gas}}=1.07\, \$\textrm{MXN}/(\textrm{L}\cdot\textrm{year})$ and $b_{\textrm{gas}}=24.30\, \$\textrm{MXN}/\textrm{L}$, according to eq.~\eqref{eq:gas}, being the coefficients for the gasoline type (91 octane) employed for new vehicles.
Data obtained from \cite{GasolinaMX}.
}
\label{fig:gas}
\end{figure}

A gasoline vehicle can travel an approximate distance $\ell_{\textrm{range}}=V_{\textrm{tank}}/C_{\textrm{gas}}$ before stopping to be refueled.
$V_{\textrm{tank}}$ is the gasoline volume that its fuel tank can store, in $\textrm{L}$, and $C_{\textrm{gas}}$ is the fuel economy or average gasoline consumption during motion, given in $\textrm{L}/\textrm{km}$.
Thus, the volume of gasoline consumed during a week $V$, given in $\textrm{L}/\textrm{week}$, can be estimated as:
\begin{eqnarray}
V(\ell,U,T) &=\ell C_{\textrm{gas}}+\left(T-\frac{\ell}{U}\right)C_{\textrm{idle}} \ ,
\label{eq:gascons}
\end{eqnarray}
where $C_{\textrm{idle}}$ is the idling consumption, in $\textrm{L}/\textrm{h}$ and $\alpha$ is a coefficient that depends on vehicular congestion, defined as $\alpha=U\, T/\ell$.
This coefficient remains in the range $\alpha\in\left[1,\infty\right)$, since $\alpha=1$ for $T=T_{\textrm{min}}$, while the value of $\alpha$ for $T=T_{\textrm{max}}$ depends on the precise values of $U$ and $\ell$.

In addition, the gasoline price is determined by the cost of crude oil price (physical, trading and financial markets), refining, distribution, marketing, and taxes. 
Examples of the complex behavior of the gasoline prices for the two different types of fuel found in Mexico City, are shown in Fig.~\ref{fig:gas}.
Nevertheless, simple trends can be obtained from historical data.
For instance, the tendency of the gasoline price $\mathcal{G}$, in $\$/\textrm{L}$, is described by the following linear function:
\begin{eqnarray}
\mathcal{G}(t)&=a_{\textrm{gas}}\left(t-y_0\right)+b_{\textrm{gas}} \ ,
\label{eq:gas}
\end{eqnarray}
where $a_{\textrm{gas}}$ is the price increase with time, given in $\$/(\textrm{L}\cdot\textrm{year})$ and $b_{\textrm{gas}}$ is the price at the beginning of $y_0$, given in $\$/\textrm{L}$.
In Fig.~\ref{fig:gas}, linear trends are also presented, which follow the general evolution of the increasing prices of gasoline.
It is important to mention that the 91 octane gasoline type is the one employed for new vehicles in Mexico.

Therefore, the total gasoline expense in $\$$ at time $t=t_0+M/12$, since the acquisition of the vehicle at time $t_0$, is given by:
\begin{eqnarray}
G(\ell,U,T,t)&=\sum_{n=0}^{M}\left[w\, V(\ell,U,T)\, \mathcal{G}\left(t_0+\frac{n}{12}\right) \right] \ .
\end{eqnarray}

\subsection{Electricity consumption and price}

\begin{figure}
\centering
\includegraphics[width=0.50\textwidth,trim=31mm 80mm 24mm 85mm,clip]{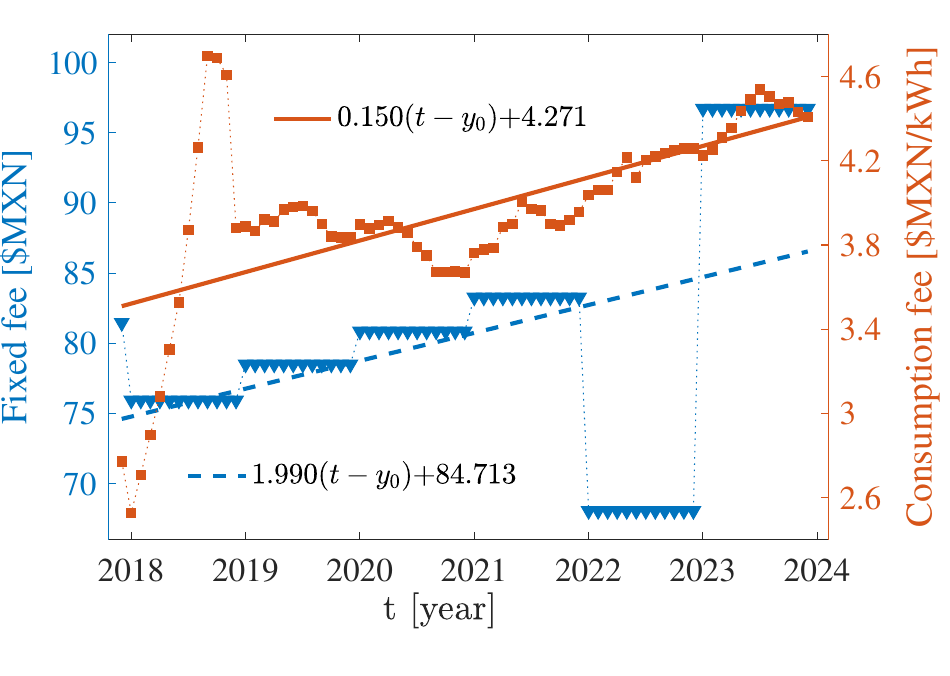}
\caption{Price of the electric energy at Mexico City. The reference year is $y_0=2023$, and linear trends are shown for the fixed fee and the consumption price, with $a_{\textrm{elec,fix}}=1.990\, \$\textrm{MXN}/\textrm{year}$, $b_{\textrm{elec,fix}}=84.713\, \$\textrm{MXN}$, $a_{\textrm{elec,cons}}=0.150\, \$\textrm{MXN}/(\textrm{kW}\cdot\textrm{h}\cdot\textrm{year})$, and $b_{\textrm{elec,cons}}=4.271\, \$\textrm{MXN}/(\textrm{kW}\cdot\textrm{h})$, according to eqs.~\eqref{eq:elec}.
Data obtained from \cite{ElectricidadMX}.
}
\label{fig:CFE}
\end{figure}

An electric vehicle can travel $\ell_{\textrm{range}}=E_{\textrm{battery}}/C_{\textrm{elec}}$ before stopping to recharge the battery.
$E_{\textrm{battery}}$ is the the energy capacity that the battery can store, in $\textrm{kW}\cdot\textrm{h}$, and $C_{\textrm{elec}}$ is the average electric energy consumption during motion, given in $(\textrm{kW}\cdot\textrm{h})/\textrm{km}$.
Thus, the energy consumed during a week $E$, given in $\textrm{kW}\cdot\textrm{h}$, can be computed as:
\begin{eqnarray}
E(\ell)&=\ell C_{\textrm{elec}} \ .
\label{eq:econs}
\end{eqnarray}

In turn, the electric energy price depends on the energy source, transformation process, and cable distribution, as well as government allowances.
Nevertheless, the behavior of the price of electric energy is much simpler than that of other energy resources, usually being organized as a fixed fee and a consumption price.
An example of the price scheme, with the fees for the electric energy that are applied in Mexico City, are presented in Fig.~\ref{fig:CFE}.

The fixed fee $F_{\textrm{elec,fix}}$, in $\$$, and the consumption price $F_{\textrm{elec,cons}}$, in $\$/(\textrm{kW}\cdot\textrm{h})$, are approximated by the following trends:
\begin{subequations}
\begin{eqnarray}
F_{\textrm{elec,fix}}(t)&=a_{\textrm{elec,fix}}\left(t-y_0\right)+b_{\textrm{elec,fix}} \ , \\
F_{\textrm{elec,cons}}(t)&=a_{\textrm{elec,cons}}\left(t-y_0\right)+b_{\textrm{elec,cons}} \ ,
\end{eqnarray}
\label{eq:elec}
\end{subequations}
where $a_{\textrm{elec,fix}}$ is the fee increase with time, given in $\$/\textrm{year}$, and $b_{\textrm{elec,fix}}$ is the fee at the beginning of $y_0$, given in $\$$.
In turn, $a_{\textrm{elec,cons}}$ is the price increase with time, given in $\$/(\textrm{kW}\cdot\textrm{h}\cdot\textrm{year})$, and $b_{\textrm{elec,cons}}$ is the price at the beginning of $y_0$, given in $\$/(\textrm{kW}\cdot\textrm{h})$.
In Fig.~\ref{fig:CFE}, exponential trends are also presented, which follow the evolution of the increasing fixed fee and consumption price.

The knowledge of the above-mentioned components allows us to compute the cost of battery recharge $\mathcal{R}$, in $\$$, as follows:
\begin{eqnarray}
\mathcal{R}(t)&=F_{\textrm{elec,fix}}(t)+w E(\ell) F_{\textrm{elec,cons}}(t) \ .
\end{eqnarray}
Therefore, the total electricity expense in $\$$ at time $t=t_0+M/12$, since the acquisition of the vehicle at time $t_0$, is given by:
\begin{eqnarray}
R(\ell,t)&=\sum_{n=0}^{M}\left[\mathcal{R}\left(t_0+\frac{n}{12}\right)\right] \ .
\end{eqnarray}

\section{Economic considerations}
\label{Sec:Eco}

\subsection{Depreciation}
\label{Subsec:Dep}

The depreciation of the vehicle starts as soon as it is acquired, leading to a decrease of its value $P$ with time.
The value of the vehicle, at year $y$, reads:
\begin{eqnarray}
P(y)&=\begin{cases}
P_0 & \textrm{for } y=y_0 \ , \\ 
\left[a_{\textrm{dep}}\left(y-y_0\right)+b_{\textrm{dep}}\right]P_0 & \textrm{for } y>y_0 \ ,
\end{cases}
\label{eq:dep}
\end{eqnarray}
where $P_0$ is the original value of the vehicle at $y_0$, given in $\$$, $a_{\textrm{dep}}$ is the fraction value reduction with passing years, in $\textrm{year}^{-1}$, and $b_{\textrm{dep}}$ is the fraction value reduction immediately after acquisition.
For the case of Mexico, the values of the coefficients $a_{\textrm{dep}}=-0.1$ $\textrm{year}^{-1}$ and $b_{\textrm{dep}}=0.83$ are considered, according to government information~\cite{DepreciacionMX}.

\subsection{Monthly percentage change in CPI}

\begin{figure}
\centering
\includegraphics[width=0.50\textwidth,trim=35mm 80mm 30mm 85mm,clip]{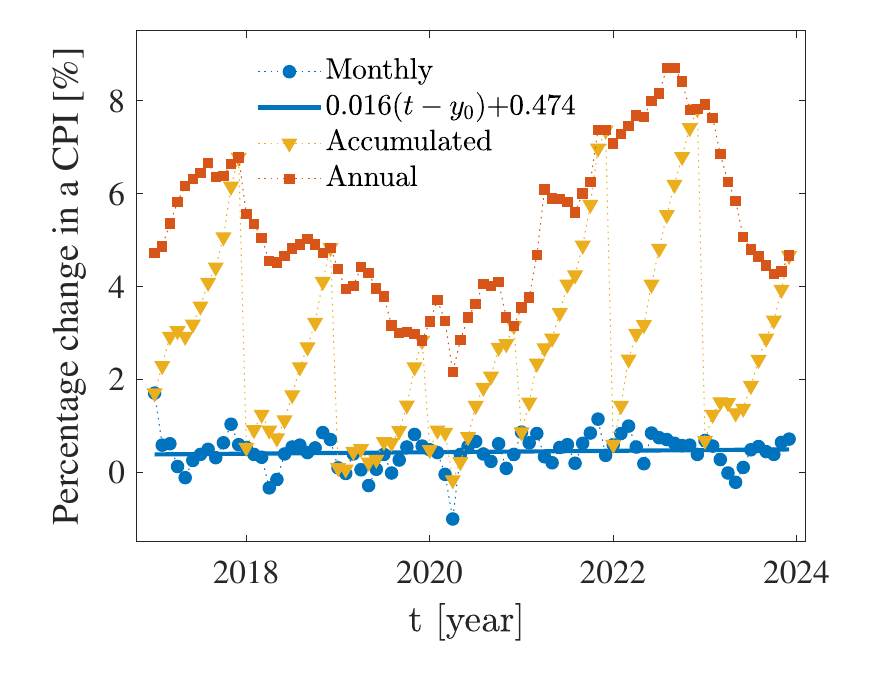}
\caption{Monthly, accumulated (from each January), and annual inflation rate in Mexico. The reference year is $y_0=2023$, and a linear trend for the monthly inflation rate $I_{\textrm{m}}$ is shown, with $a_{\textrm{m}}=0.016\, \%/\textrm{year}$ and $b_{\textrm{m}}=0.474\, \%$, according to eq.~\eqref{eq:infla}.
Data obtained from \cite{InflationMX}.
}
\label{fig:infla}
\end{figure}

The monthly percentage change in CPI (Consumer Price Index) or inflation rate $I_{\textrm{m}}$ is approximated by:
\begin{eqnarray}
I_{\textrm{m}}(t)&=a_{\textrm{m}}\left(t-y_0\right)+b_{\textrm{m}} \ ,
\label{eq:infla}
\end{eqnarray}
where $a_{\textrm{m}}$ is the inflation rate increase with time, given in $\%/\textrm{year}$ and $b_{\textrm{m}}$ is the inflation rate at the beginning of $y_0$, in $\%$.
In Fig.~\ref{fig:infla}, historical data of the monthly inflation rate for Mexico is shown, together with the accumulated rate, from the beginning of each year, and the annual rate, which corresponds to the accumulated value for one year, starting at the same month from the preceding year.
In contrast with the accumulated and annual behaviors, the monthly inflation rate displays a more stable evolution. 

Thence, due to the nature of the monthly inflation rate $I_{\textrm{m}}$, a prediction of its evolution over time allows to estimate the future cost of goods and services.
If the cost of a replacement part or a service $Q_0$ is given at $t_0$, the price $Q$ at time $t=t_0+M/12$ will be given by:
\begin{eqnarray}
Q(t)&=Q_0 \prod_{n=1}^{M}\left\{1+\frac{1}{100}\left[I_{\textrm{m}}\left(t_0+\frac{n}{12}\right)\right]\right\} \ .
\label{eq:incP}
\end{eqnarray}

\subsection{Maintenance and technical services}
\label{Subsec:Tech}

According to manufacturers' advice~\cite{NissanMX,JacMX}, the maintenance and technical services for a new vehicle should be scheduled every $10000$ km or 6 months, the one occurring the first.
Additionally, they do not have the same cost, which also evolves in time, even though a value chart, corresponding only to time $t_0$, is presented when the vehicle is acquired.
For vehicles that are meant for weekly driving distances shorter than $\ell\approx(10000 \, \textrm{km}/6w\, \textrm{week})\approx 383.6 \, \textrm{km}/\textrm{week}$, the maintenance and technical services must be performed each 6 months and with a price that evolves according to the monthly inflation rate.
Therefore, using eq.~\eqref{eq:incP}, at time $t=t_0+M/12$ the total amount payed for these services should be given by:
\begin{subequations}
\begin{eqnarray}
S(t)&=\sum_{n=1}^M \left(P_0\left[\vphantom{\frac{a}{a}}\Psi_1(n)+\Psi_2(n)+\Psi_3(n)\right]\times\right. \notag \\
& \hspace*{20mm} \left.\prod_{q=1}^{n}\left\{1+\frac{1}{100}\left[I_{\textrm{m}}\left(t_0+\frac{q}{12}\right)\right]\right\}\right) \ ,
\end{eqnarray}
with:
\begin{eqnarray}
\Psi_1(n)&=
\begin{cases}
\psi_1 & \textrm{if } n/6\in\mathbb{Z}^{+} \ , \\
0 & \textrm{otherwise} \ ,
\end{cases} \notag \\
\Psi_2(n)&=
\begin{cases}
\psi_2 & \textrm{if } n/12\in\mathbb{Z}^{+} \ , \\
0 & \textrm{otherwise} \ ,
\end{cases} \notag \\
\Psi_3(n)&=
\begin{cases}
\psi_3 & \textrm{if } n/24\in\mathbb{Z}^{+} \ , \\
0 & \textrm{otherwise} \ ,
\end{cases}
\end{eqnarray}
\end{subequations}
being $\psi_1$, $\psi_2$, and $\psi_3$ the fractions of $P_0$ that are considered for the cost of maintenance and technical services, given at time $t_0$, with frequencies of $(6 \, \textrm{month})^{-1}$, $(12 \, \textrm{month})^{-1}$, and $(24 \, \textrm{month})^{-1}$, respectively.
The values of $\psi_1$, $\psi_2$, and $\psi_3$ are calculated according to information provided by the manufacturers.
In table~\ref{tab:params}, the values for the vehicles studied in this work are presented.

\subsection{Taxation}

\begin{figure}
\centering
\includegraphics[width=0.50\textwidth,trim=32mm 105mm 33mm 45mm,clip]{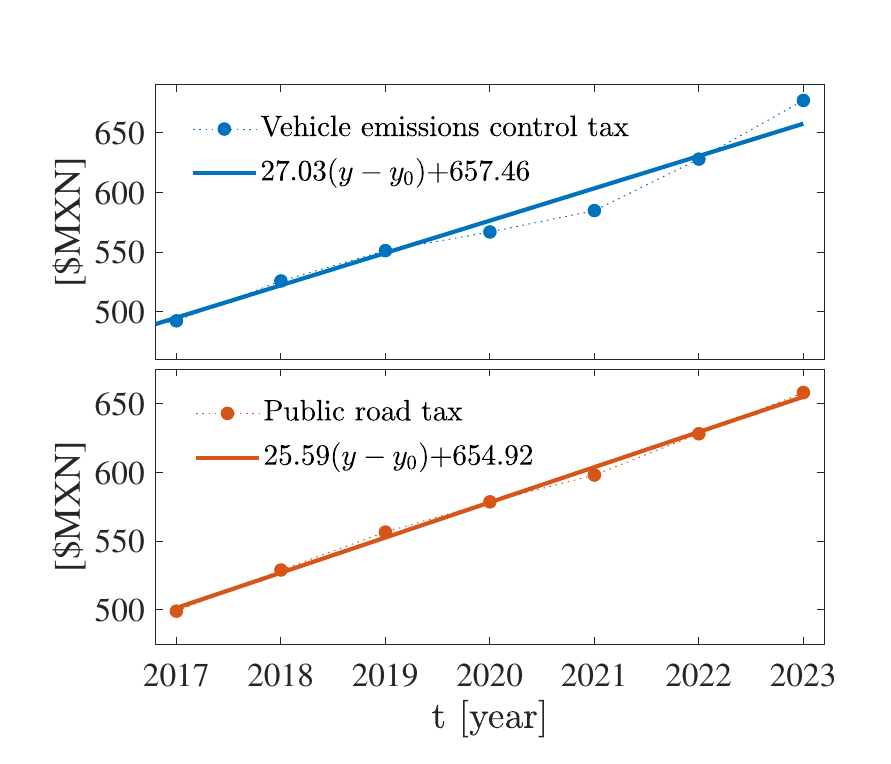}
\caption{Vehicle emission control and public road taxes in Mexico City. The reference year is $y_0=2023$, and linear trends are shown for each tax, with $a_{\textrm{vec}}=27.03\, \$\textrm{MXN}/\textrm{year}$, $b_{\textrm{vec}}=657.46\, \$\textrm{MXN}$, $a_{\textrm{road}}=25.59\, \$\textrm{MXN}/\textrm{year}$, and $b_{\textrm{road}}=654.92\, \$\textrm{MXN}$, according to eqs.~\eqref{eq:vec} and \eqref{eq:road}, respectively.
Data obtained from \cite{UMAMX} and \cite{TenenciaMX}.}
\label{fig:taxes}
\end{figure}

Depending on the specific location (country, region and city), different taxes may apply, which must be paid to allow the use of the vehicle.
Among the common taxes, we find the vehicle emission control tax, the public road tax and the ownership tax.
In Fig.~\ref{fig:taxes}, historical data of the vehicle emission control tax and the public road tax, that are applied in Mexico City, are presented.

The vehicle emissions control tax $A_{\textrm{vec}}$ is approximated by:
\begin{eqnarray}
A_{\textrm{vec}}(y)&=f_{\textrm{vec}}\left[a_{\textrm{vec}}\left(y-y_0\right)+b_{\textrm{vec}}\right] \ ,
\label{eq:vec}
\end{eqnarray}
where $f_{\textrm{vec}}$ is the frequency of the controls in a year, $a_{\textrm{vec}}$ is the tax increase with time, given in $\$/\textrm{year}$ and $b_{\textrm{gas}}$ is the tax at $y_0$, in $\$$.
For instance, in Mexico the emission control tax is given by $c_{\textrm{vec}}(1+p_{\textrm{vat}}) \textrm{UMA}$~\cite{UMAMX,TenenciaMX}, where 
$p_{\textrm{vat}}=0.16$ is the value-added tax and $c_{\textrm{vec}}=5.625$ is a coefficient fixed by the government, whereas the time variation is given by UMA, the Unit of Measure and Update (minimum wage per day).
A comparison of this information with eq.~\eqref{eq:vec} yields the values of the coefficients $a_{\textrm{vec}}$ and $b_{\textrm{vec}}$ that apply for Mexico City.

The public road tax $A_{\textrm{road}}$ is estimated by:
\begin{eqnarray}
A_{\textrm{road}}(y)&=f_{\textrm{road}}\left[a_{\textrm{road}}\left(y-y_0\right)+b_{\textrm{road}}\right] \ ,
\label{eq:road}
\end{eqnarray}
where $f_{\textrm{road}}$ is the frequency of tax imposition in a year, $a_{\textrm{road}}$ is the tax increase with time, given in $\$/\textrm{year}$ and $b_{\textrm{road}}$ is the tax at $y_0$, in $\$$.

The ownership tax $A_{\textrm{own}}$ is calculated as:
\begin{eqnarray}
A_{\textrm{own}}(y)&=\begin{cases}
0 & \textrm{if } P(y)<P_{\textrm{th}} \ , \\
f_{\textrm{own}}b_{\textrm{own}} P(y) & \textrm{if } P(y)\geq P_{\textrm{th}} \ ,
\end{cases}
\label{eq:own}
\end{eqnarray}
where $P$ is the value of the vehicle at year $y$, given by eq.~\eqref{eq:dep}, $f_{\textrm{own}}$ is the frequency of tax imposition in a year, $b_{\textrm{own}}$ is the tax coefficient, and $P_{\textrm{th}}$ is a threshold value for ownership tax imposition.
For Mexico City, the trends and values of the coefficients for the vehicle emissions control and the public road taxes are given in Fig.~\ref{fig:taxes}, while $P_{\textrm{th}}=250000\, \$\textrm{MXN}$, according to governmental information~\cite{TenenciaMX}.

Therefore, the total taxation, from $t_0$ to $t$, is:
\begin{eqnarray}
A(t)&=\sum_{n=0}^Y \left[A_{\textrm{vec}}(y_0+n)+A_{\textrm{road}}(y_0+n)+A_{\textrm{own}}(y_0+n)\right] \ ,
\end{eqnarray}

Since electric vehicles do not cause emissions, it is likely to find $A_{\textrm{vec}}=0$ for them.
Additionally, many countries apply tax incentives for the acquisition of electric vehicles, for instance, reducing $A_{\textrm{own}}$ and $A_{\textrm{road}}$ throughout the life cycle of the electric vehicle. 
For the case of Mexico City, $A_{\textrm{own}}=0$ and $A_{\textrm{road}}$ is multiplied by a $1/2$ coefficient~\cite{EVMX}.

\section{Total expenses}
\label{Sec:Total}

As should be expected, acquiring a vehicle entails financial consequences, which one must evaluate and weight, according to the personal economical situation, in order to make an adequate choice.
The main objective of this study is to provide a tool to perform such a task.
For this reason, all the information, presented in Sections~\ref{Sec:Motion} and \ref{Sec:Eco} of the present manuscript, has been gathered and included in a simple model, which purpose is to yield the total of significant operating costs, for gasoline and electric vehicles.
Two alternatives are proposed in order to compare the expenditure of an electric vehicle against a gasoline one over time.
As a first alternative, the owner is considered to keep the vehicle for its entire lifespan, whereas as a second approach, the owner may recover a partial amount of the vehicle value by selling it, after a certain time.
In the following, both options are described for each type of vehicle.

\subsection{Gasoline vehicle}
\label{Subsec:Gas}

The total expense $X_{\textrm{gas}}$ that the gasoline vehicle owner will have spent at time $t$, since the acquisition of the vehicle at time $t_0$, reads:
\begin{subequations}
\begin{eqnarray}
X_{\textrm{gas}}(\ell,U,T,t)&=P_{\textrm{gas},0}+G(\ell,U,T,t)+S_{\textrm{gas}}(t)+A_{\textrm{gas}}(t) \ ,
\label{eq:gasexp}
\end{eqnarray}
whereas the total loss is:
\begin{eqnarray}
Z_{\textrm{gas}}(\ell,U,T,t)&=X_{\textrm{gas}}(\ell,U,T,t)-P_{\textrm{gas}}(t) \ .
\label{eq:gasloss}
\end{eqnarray}
\end{subequations}
which is given by the difference between the total expense $X_{\textrm{gas}}$, at time $t$, and the amount $P_{\textrm{gas}}$ that would be recovered by selling the vehicle, also at $t$.

\subsection{Electric vehicle}
\label{Subsec:Elec}

The total expense $X_{\textrm{elec}}$ that the electric vehicle owner will have spent at time $t$, since the acquisition of the vehicle at time $t_0$, is given by:
\begin{subequations}
\begin{eqnarray}
X_{\textrm{elec}}(\ell,t)&=P_{\textrm{elec},0}+R(\ell,t)+S_{\textrm{elec}}(t)+A_{\textrm{elec}}(t) \ ,
\label{eq:elecexp}
\end{eqnarray}
whereas the total loss is:
\begin{eqnarray}
Z_{\textrm{elec}}(\ell,t)&=X_{\textrm{elec}}(\ell,t)-P_{\textrm{elec}}(t) \ ,
\label{eq:elecloss}
\end{eqnarray}
\end{subequations}
which is given by the difference between the total expense $X_{\textrm{elec}}$, at time $t$, and the amount $P_{\textrm{elec}}$ that would be recovered by selling the vehicle, also at $t$.

\section{Numeric calculations}
\label{Subsec:AFW}

The following equations correspond to algebraic simplifications of the original expressions introduced in Sections~\ref{Sec:Motion}--\ref{Sec:Total}.
They were employed to perform the calculations and obtain the results presented in Section~\ref{Sec:Res}.

For given values of $\ell$, $U$ and $\alpha$, the volume of gasoline consumed during a week is computed only once as follows:
\begin{eqnarray}
V&=\ell C_{\textrm{gas}}\left[1+\left(\alpha-1\right)\frac{C_{\textrm{idle}}}{U C_{\textrm{gas}}}\right] \ .
\label{eq:vlua}
\end{eqnarray}
Considering the acquisition time $t_0=y_0+m_0/12$, one finds for month $m$ of year $y$:
\begin{eqnarray}
M&=12(y-y_0)+m-m_0 \notag \\
Y&=\Bigg\lfloor y-y_0+\dfrac{m-m_0}{12}\Bigg\rfloor \ .
\end{eqnarray}
Now, the value of $V$, obtained with eq.~\eqref{eq:vlua}, is employed to calculate the gasoline expense at month $m$ of year $y$ as:
\begin{eqnarray}
G(m,y)
&=w \left\{\frac{a_{\textrm{gas}}}{12}\left[m_0+\frac{M}{2}\right]+b_{\textrm{gas}}\right\} \left[M+1\right] V \ .
\end{eqnarray}
As well, the electricity accumulated cost at $m$ and $y$ is:
\begin{eqnarray}
R(m,y)
&=
\left\{
w E(\ell)\left(\frac{a_{\textrm{elec,cons}}}{12}\left[m_0+\frac{M}{2}\right]+b_{\textrm{elec,cons}}\right)
\right. \notag \\
&\hspace*{30mm}
\left.
+\frac{a_{\textrm{elec,fix}}}{12}\left[m_0+\frac{M}{2}\right]+b_{\textrm{elec,fix}}
\right\} \left[M+1\right]
\end{eqnarray}

The maintenance and technical services expense covered at month $m$ of year $y$ is given by:
\begin{eqnarray}
S(m,y)&=P_0\left[
\Psi_1\sum_{n=1}^{\lfloor M/6\rfloor} \left(\prod_{q=1}^{6n}\left\{1+\frac{1}{100}\left[\frac{a_{\textrm{m}}\left(m_0+q\right)}{12}+b_{\textrm{m}}\right]\right\}\right)
\right. \\
& \hspace*{15mm}
+\Psi_2\sum_{n=1}^{\lfloor M/12\rfloor} \left(\prod_{q=1}^{12n}\left\{1+\frac{1}{100}\left[\frac{a_{\textrm{m}}\left(m_0+q\right)}{12}+b_{\textrm{m}}\right]\right\}\right) \notag \\
& \hspace*{20mm}
\left.
+\Psi_3\sum_{n=1}^{\lfloor M/24\rfloor} \left(\prod_{q=1}^{24n}\left\{1+\frac{1}{100}\left[\frac{a_{\textrm{m}}\left(m_0+q\right)}{12}+b_{\textrm{m}}\right]\right\}\right)
\right]\ . \notag
\end{eqnarray}

In turn, the depreciated value of the vehicle is computed with eq.~\eqref{eq:dep}, whereas the accumulated taxes, at month $m$ of year $y$, are calculated as follows:
\begin{eqnarray}
A(m,y)
&=A_{\textrm{vr}}(m,y)+A_{\textrm{o}}(m,y) \ , \\
A_{\textrm{vr}}(m,y)&=(Y+1)
\left[\left(\frac{f_{\textrm{vec}}a_{\textrm{vec}}+f_{\textrm{road}}a_{\textrm{road}}}{2}\right)Y \right. \notag \\
& \hspace*{10mm}
\left.
+\left(f_{\textrm{vec}}b_{\textrm{vec}}+f_{\textrm{road}}b_{\textrm{road}}\right)\right] \ , \notag \\
A_{\textrm{o}}(m,y)&=\begin{cases}
0 & \textrm{if } P_0<P_{\textrm{th}} \ , \\
f_{\textrm{own}}b_{\textrm{own}}P_0\times & \\
\hspace*{5mm}
\left\{1+Y_{\textrm{dep}}\left[\left(\dfrac{a_{\textrm{dep}}}{2}\right)(Y_{\textrm{dep}}+1)+b_{\textrm{dep}}\right]\right\} & \textrm{if } P_0\geq P_{\textrm{th}} \ , \notag
\end{cases}
\end{eqnarray}
where $Y_{\textrm{dep}}$ is the number of years that the depreciated value of the car is still above the threshold value $P_{\textrm{th}}$, which is given by:
\begin{eqnarray}
Y_{\textrm{dep}}&=\begin{cases}
0 & \textrm{if } P_0\leq P_{\textrm{th}}/b_{\textrm{dep}} \ , \\
\left\lfloor\dfrac{1}{a_{\textrm{dep}}}\left(\dfrac{P_{\textrm{th}}}{P_0}-b_{\textrm{dep}}\right)\right\rfloor & \textrm{otherwise} \ ,
\end{cases}
\end{eqnarray}

Finally, the total expenses and losses for the gasoline vehicle are computed as:
\begin{eqnarray}
X_{\textrm{gas}}(m,y)&=P_{\textrm{gas},0}+G(m,y)+S_{\textrm{gas}}(m,y)+A_{\textrm{gas}}(m,y) \ , \notag \\
Z_{\textrm{gas}}(m,y)&=X_{\textrm{gas}}(m,y)-P_{\textrm{gas}}(y) \ ,
\end{eqnarray}
whereas for the electric vehicle, the total expenses and losses are obtained with:
\begin{eqnarray}
X_{\textrm{elec}}(m,y)&=P_{\textrm{elec},0}+R(m,y)+S_{\textrm{elec}}(m,y)+A_{\textrm{elec}}(m,y) \ , \notag \\    
Z_{\textrm{elec}}(m,y)&=X_{\textrm{elec}}(m,y)-P_{\textrm{elec}}(y) \ .
\end{eqnarray}

A Matlab code was developed to perform the calculations, which can be shared on demand, by contacting the authors.

\section{Results: Specific comparison of vehicles at Mexico city}
\label{Sec:Res}

\begin{table*}[t]
\centering
\caption{Characteristics of the less expensive gasoline and electric  vehicles in Mexico.
Data obtained from the public information distributed by the manufacturers~\cite{NissanMX, JacMX}.
Notice that $\psi_2$ is negative for the electric vehicle example, since the total cost of the services (equal to $P_0\left[\psi_1+\psi_2+\psi_3\right]$), at even multiples of 6 months, is lower than those at odd multiples.}
\begin{tabular}{|>{\centering\arraybackslash}p{2.5cm}|*{2}{>{\centering\arraybackslash}p{5.5cm}|}}
\hline
Vehicle type & Gasoline & Electric \\ \hline
Manufacturer & Nissan~\cite{NissanMX} & Jac~\cite{JacMX} \\
& \footnotesize Nissan Motor Co. Ltd. & \footnotesize Anhui Jianghuai Automobile Co., Ltd. \\ \hline
Model name & March (Micra) &  E10x \\ \hline
$P_0$ & $257900.00$ $\$$MXN & $439000.00$ $\$$MXN \\ \hline
$\psi_1$ & $7.968\times10^{-3}$ & $3.506\times10^{-3}$ \\ \hline
$\psi_2$ & $2.675\times10^{-3}$ & $-1.640\times10^{-3}$ \\ \hline
$\psi_3$ & $3.742\times10^{-3}$ & $4.100\times10^{-4}$ \\ \hline
$C_{\textrm{gas}}$ & $1/16$ L$/$km & Not applicable \\ \hline
$C_{\textrm{idle}}$ & $1/2$ L$/$km & Not applicable \\ \hline
$E_{\textrm{battery}}$ & Not applicable & $31.4$ (kW$\cdot$h) \\ \hline
$C_{\textrm{elec}}$ & Not applicable & $0.1040$ (kW$\cdot$h)$/$km \\ \hline
\end{tabular}
\label{tab:params}
\end{table*}

\begin{table*}[t]
\centering
\caption{Values of the coefficients employed for the calculation of the total expenses and losses of the gasoline and electric vehicles in Mexico City.
Data obtained from the public information provided by several government departments~\cite{GasolinaMX,ElectricidadMX,DepreciacionMX,InflationMX,UMAMX,TenenciaMX,EVMX}.}
\begin{tabular}{|>{\centering\arraybackslash}p{2.5cm}|*{2}{>{\centering\arraybackslash}p{5.5cm}|}}
\hline
Coefficient & Gasoline & Electric \\ \hline
$a_{\textrm{gas}}$ & 1.069 $\$\textrm{MXN}/(\textrm{L}\cdot\textrm{year})$ & Not applicable \\ \hline
$b_{\textrm{gas}}$ & 24.300 $\$\textrm{MXN}/(\textrm{L})$ & Not applicable \\ \hline
$a_{\textrm{elec,fix}}$ & Not applicable & 1.990 $\$\textrm{MXN}/\textrm{year}$ \\ \hline
$b_{\textrm{elec,fix}}$ & Not applicable & 84.713 $\$\textrm{MXN}$ \\ \hline
$a_{\textrm{elec,cons}}$ & Not applicable & 0.150 $\$\textrm{MXN}/(\textrm{kW}\cdot\textrm{h}\cdot\textrm{year}$ \\ \hline
$b_{\textrm{elec,cons}}$ & Not applicable & 4.271 $\$\textrm{MXN}/(\textrm{kW}\cdot\textrm{h}$ \\ \hline
$a_{\textrm{dep}}$ & \multicolumn{2}{c|}{-0.100 $\textrm{year}^{-1}$} \\ \hline
$b_{\textrm{dep}}$ & \multicolumn{2}{c|}{0.830} \\ \hline
$a_{\textrm{m}}$ & \multicolumn{2}{c|}{0.016 $\%/\textrm{year}$} \\ \hline
$b_{\textrm{m}}$ & \multicolumn{2}{c|}{0.474 $\%$} \\ \hline
$a_{\textrm{vec}}$ & 27.03 $\$\textrm{MXN}/\textrm{year}$ & Not applicable \\ \hline
$b_{\textrm{vec}}$ & 657.46 $\$\textrm{MXN}$ & Not applicable \\ \hline
$a_{\textrm{road}}$ & 25.59 $\$\textrm{MXN}/\textrm{year}$ & 12.80 $\$\textrm{MXN}/\textrm{year}$ \\ \hline
$b_{\textrm{road}}$ & 654.92 $\$\textrm{MXN}$ & 327.46 $\$\textrm{MXN}$ \\ \hline
$f_{\textrm{own}}$ & 1 & Not applicable \\ \hline
$b_{\textrm{own}}$ & 0.030 & Not applicable \\ \hline
$P_{\textrm{th}}$ & 250000 $\$\textrm{MXN}$ & Not applicable \\ \hline
\end{tabular}
\label{tab:coeffs}
\end{table*}

Applying the proposed methodology, a comparison between a conventional gasoline vehicle and an electric vehicle, whose characteristics are described in Table~\ref{tab:params}, is presented within the context of Mexico City.
Additionally, the values of the coefficients of the trends, employed for the calculations of the corresponding expenses, are given in Table~\ref{tab:coeffs}.
It is also worth noting that this study is restricted to personal use vehicles, for which the aforementioned considerations are valid.
Commercial purpose automobiles, such as taxis or delivery transports, which usually cover significantly greater distances compared to personal use vehicles, are not included in the following analysis.
Nevertheless, the presented methodology can be adapted to that case, by modifying the frequency of maintenance and technical services, described in Section~\ref{Subsec:Tech}.

\begin{figure*}[t]
\centering
\includegraphics[width=0.75\textwidth]{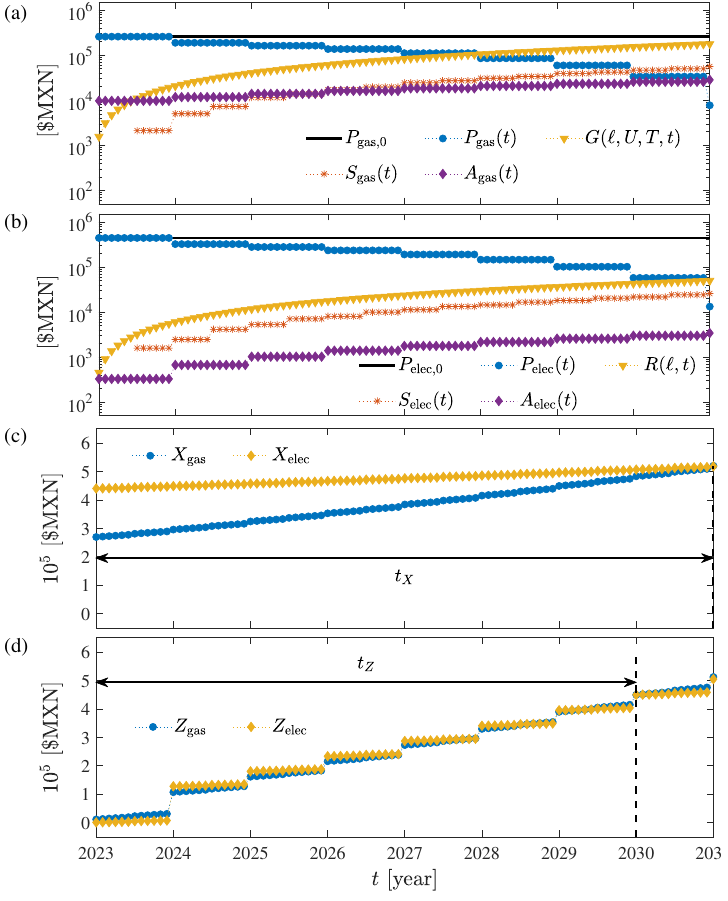}
\caption{Evolution with time $t$ of prices, at acquisition $P_0$ and depreciated $P$, and expenses, of gasoline $G$, electric energy $R$, services $S$ and taxes $A$, for (a) the gasoline vehicle and (b) the electric vehicle, which details are given in Table~\ref{tab:params}.
Evolution with time $t$ of (c) the total expenses $X$ and (d) the total loss $Z$, of the gasoline and electric vehicles.
The equilibrium times, for which the expenses and losses for both vehicles are equal, are also shown. 
The data has been computed for $\ell=190\, \textrm{km}/\textrm{week}$, $U=25\, \textrm{km}/\textrm{h}$ and $T=\alpha\ell/U$, with $\alpha=1.75$, according to eqs.~\eqref{eq:gascons} and \eqref{eq:econs}.}
\label{fig:ell190}
\end{figure*}

\begin{figure*}[t]
\centering
\includegraphics[width=0.75\textwidth]{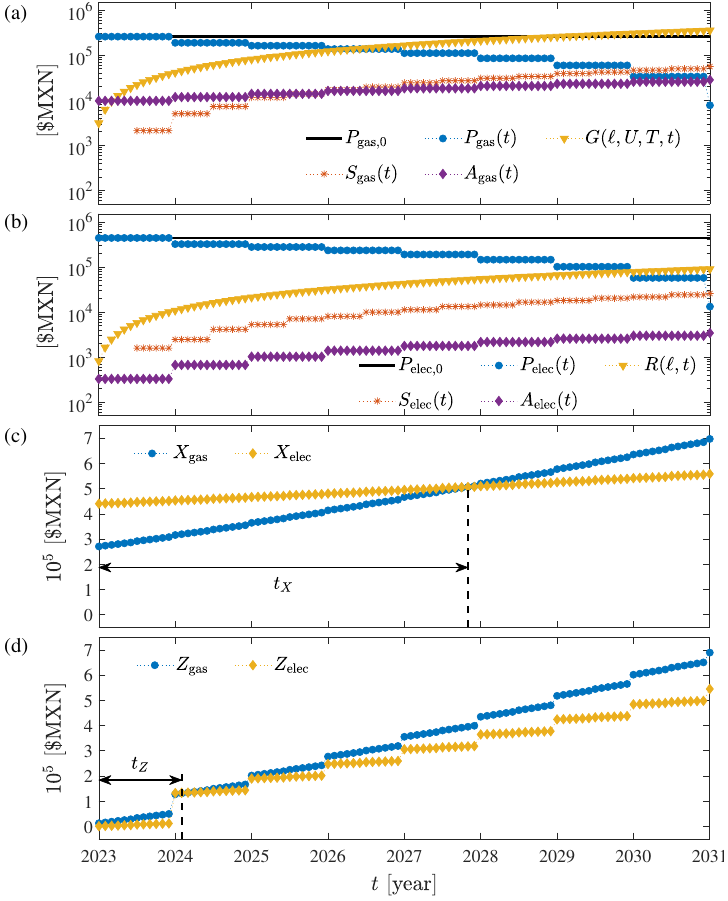}
\caption{Evolution with time $t$ of prices, at acquisition $P_0$ and depreciated $P$, and expenses, of gasoline $G$, electric energy $R$, services $S$ and taxes $A$, for (a) the gasoline vehicle and (b) the electric vehicle, which details are given in Table~\ref{tab:params}.
Evolution with time $t$ of (c) the total expenses $X$ and (d) the total loss $Z$, of the gasoline and electric vehicles.
The equilibrium times, for which the expenses and losses for both vehicles are equal, are also shown. 
The data has been computed for $\ell=380\, \textrm{km}/\textrm{week}$, $U=25\, \textrm{km}/\textrm{h}$ and $T=\alpha\ell/U$, with $\alpha=1.75$, according to eqs.~\eqref{eq:gascons} and \eqref{eq:econs}.}
\label{fig:ell380}
\end{figure*}

\begin{figure*}[t]
\centering
\includegraphics[width=0.75\textwidth]{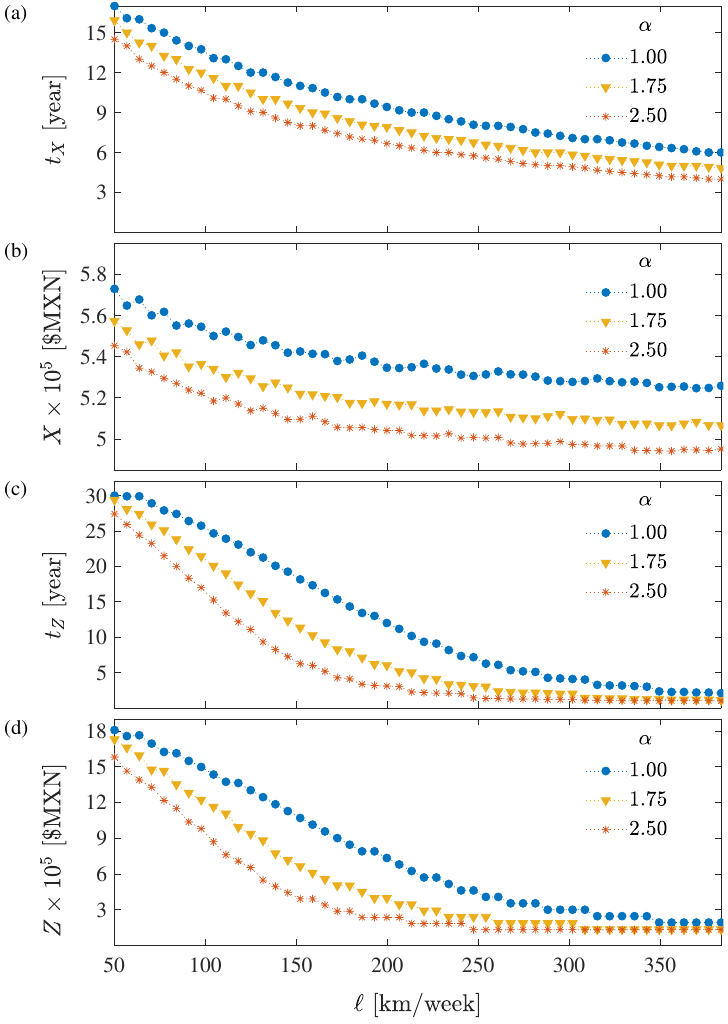}
\caption{Dependence on the weekly driving distance $\ell$, of (a) the time for equal expenses $t_X$, (b) the equilibrium expenses $X$, (c) the time for equal losses $t_Z$, (b) the equilibrium losses $X$, when comparing the gasoline and electric vehicles, which details are given in Table~\ref{tab:params}.
The data has been computed for $U=25\, \textrm{km}/\textrm{h}$ and $T=\alpha\ell/U$, with three different values of $\alpha$, according to eqs.~\eqref{eq:gascons} and \eqref{eq:econs}.
}
\label{fig:elldependence}
\end{figure*}

According to previous studies and surveys developed for the U.S.~\cite{Simeu2018}, the driving distance can be described by a gamma distribution with a mean of $434 \, \textrm{km}/\textrm{week}$ and a mode of $140\, \textrm{km}/\textrm{week}$.
In this work, we decided to sweep the range $[50,380]\, \textrm{km}/\textrm{week}$, the upper limit being slightly below the driving distance of $383.6\, \textrm{km}/\textrm{week}$, which corresponds to the threshold, advised by manufacturers, for the change of the frequency of maintenance and services.
As example results, Fig.~\ref{fig:ell190} and Fig.~\ref{fig:ell380} show the expenses as functions of time, for the gasoline and electric vehicles, close to the middle and upper limit of the tested range of $\ell$, respectively.
Fig.~\ref{fig:elldependence} summarizes the equilibrium results (expenses, losses and times for equal values), from which the economic advantages of the electric vehicle  overcome those of the gasoline vehicle, as functions of the driving distance per week $\ell$.

Following eqs.~\eqref{eq:gascons} and \eqref{eq:econs}, respectively for the gasoline and the electric energy consumption, the values of several parameters must be defined, in addition to the ones enlisted in Table~\ref{tab:params}, in order to evaluate and compare their economic performance. 
For instance, a driving distance of $\ell=190\, \textrm{km}/\textrm{week}$ was considered for computing the gasoline expense $G$ and the electricity expense $R$ presented in Figs.~\ref{fig:ell190}a and \ref{fig:ell190}b.
This distance is equivalent to $4940\, \textrm{km}$ in 6 months, which is slightly below the midpoint mileage ($5000\, \textrm{km}$), according to the standard schedule of maintenance and technical services detailed in Section~\ref{Subsec:Tech}, indicating low vehicle usage.
Now, taking into account that the speed limit in most roads at Mexico City is $50\, \textrm{km}/\textrm{h}$, an average speed $U=25\, \textrm{km}/\textrm{h}$ has been proposed for the gasoline vehicle evaluation.
Traffic conditions in Mexico City usually lead to $\alpha>1$, or equivalently to $T>T_{min}$, with $T_{\textrm{min}}=\ell/U$ as defined in Section~\ref{Sec:Motion}.
Thus, $\alpha=1.75$ was employed to obtain the results depicted for the gasoline vehicle in Figure~\ref{fig:ell190}a.
This value has been selected from personal experience, in order to show characteristic results for Mexico City.
Nevertheless, the effect of $\alpha$ has also been explored in this work, with the corresponding main results being presented in Fig.~\ref{fig:elldependence}.

Figures~\ref{fig:ell190}a and \ref{fig:ell190}b display the acquisition prices of the vehicles, $P_{\textrm{gas},0}$ and $P_{\textrm{elec},0}$, and their corresponding depreciation, $P_{\textrm{gas}}$ and $P_{\textrm{elec}}$, over the years and calculated since its acquisition.
Both depreciation values decrease in steps of a yearly width, since that is the way manufacturers and the corresponding government departments agreed, as explained in Section.~\ref{Subsec:Dep}. 
It also shows fuel $G$ and electric energy $R$ expenses, considering their increasing evolutions with time.
These expenses present a linearly increasing behavior, as depicted in Figs.~\ref{fig:gas} and \ref{fig:CFE}
As well, the maintenance and technical services costs $S_{\textrm{gas}}$ and $S_{\textrm{elec}}$ are also depicted, taking into account the inflation trend of Mexico.
Once more, a stepped behavior is observed, since the expenses of the services occurs periodically (6 months), maintaining a constant accumulated value until the next service is performed.
Finally, Fig.~\ref{fig:ell190}a includes the addition of emissions control, public road and ownership taxes (when they apply), represented by the variables $A_{\textrm{gas}}$, for the gasoline vehicle, and $A_{\textrm{elec}}$ for the electric vehicle.
Taxes are also applied periodically once a year, thus a staggered curve is once again discerned.
Notice that only the acquisition price and depreciation of the electric vehicle shows higher levels than those of the gasoline vehicle, whereas the other variables (energy, services and taxes) of the electric vehicle remain below the equivalent magnitudes of the gasoline vehicles.

As described in eqs.~\eqref{eq:gasexp} and \eqref{eq:elecexp}, the addition of the terms shown in Figs.~\ref{fig:ell190}a and \ref{fig:ell190}b, excluding the depreciation of the vehicles, leads to the determination of the total expenses $X_{\textrm{gas}}$ and $X_{\textrm{elec}}$, respectively for the gasoline and electric vehicles.
These expenses are presented as functions of time $t$, and compared to each other in Fig.~\ref{fig:ell190}c, in which both show an almost linear growing behavior.
The total gasoline expenses $X_{\textrm{gas}}$ displays a lower value than the total electric expenses $X_{\textrm{elec}}$, during more than 7 years.
The time at which $X_{\textrm{elec}}$ becomes equal to $X_{\textrm{gas}}$ is marked as $t_X\approx 7.8\, \textrm{years}$, with respect to the acquisition of the vehicle.
For $t>t_X$, the value of $X_{\textrm{gas}}$ turns out to be larger than $X_{\textrm{elec}}$.
Therefore, considering that the owner keeps the vehicle during its entire lifespan, savings for the electric vehicle are expected to be observed for $t\geq t_X$.
For $t<t_X$, the gasoline vehicle entails lower expenses.
The equilibrium expense that occurs at $t_X$ is $X\approx 5.1\times10^5\, \$\textrm{MXN}$.

The owner can recover the current or depreciated value of the vehicle, by selling it at any time.
According to eqs.~\eqref{eq:gasloss} and \eqref{eq:elecloss}, by subtracting the depreciated price of the corresponding vehicle to the total expenses at a given time, the total losses $Z_{\textrm{gas}}$ and $Z_{\textrm{elec}}$ are obtained, respectively for the gasoline and electric vehicles.
These losses are presented as functions of time $t$, and compared to each other in Fig.~\ref{fig:ell190}d.
Even though both of them present a monotonic growing evolution, a step-like sudden increase is observed at the beginning of each year, with different increments for each year and for each vehicle.
The staggered evolution is a consequence of the stepped behavior of the depreciation and the costs of services and taxes, whereas the main component of the monotonic growing is due to the expenses on fuel and electric energy.  
The intersection of the gasoline loss with the electric loss each year occurs at $t_Z=7\, \textrm{years}$.
For $t<t_Z$, $Z_{\textrm{gas}}$ remains lower than $Z_{\textrm{elec}}$ at least for a fraction of each year, whereas for $t>t_Z$, $Z_{\textrm{gas}}$ is larger than $Z_{\textrm{elec}}$ for all the year.
Therefore, the electric vehicle loss becomes definitively lower than the gasoline vehicle loss, if the vehicle is sold at any $t\geq t_Z$.
Selling the vehicle before $t=t_Z$ implies a slight advantage for the gasoline vehicle.
The equilibrium loss that occurs at $t_Z$ is $Z\approx 4.5\times10^5\, \$\textrm{MXN}$.

Now, for the same average speed $U=25\, \textrm{km}/\textrm{h}$ and the same time spent in the vehicle $T=\alpha\ell/U$, with $\alpha=1.75$, a high vehicle usage is considered by taking a driving distance of $\ell=380\, \textrm{km}/\textrm{week}$.
This value of $\ell$ corresponds to $9880\, \textrm{km}$ in 6 months, which is slightly below the endpoint mileage ($10000\, \textrm{km}$), according to the standard schedule of maintenance and technical services detailed in Section~\ref{Subsec:Tech}.
The curves in Fig.~\ref{fig:ell380}, that represent the acquisition price, the depreciation values, maintenance and services, and taxes, are the same as for Fig.~\ref{fig:ell190}, and are shown for comparison purposes.
Energy consumption expenses for gasoline $G$ and for electricity expenses $R$ have been calculated for the driving distance $\ell=380\, \textrm{km}/\textrm{week}$, which results are depicted in Figs.~\ref{fig:ell380}a and \ref{fig:ell380}b, respectively.
They exhibit a similar behavior to the trends presented in Fig.~\ref{fig:ell190}, but with significant increases in magnitude. 
These changes produce a reduction of $t_X$ from 7.9 years, for $\ell=190\, \textrm{km}/\textrm{week}$, to 5.1 years, for $\ell=380\, \textrm{km}/\textrm{week}$, as it can be discerned from Fig.~\ref{fig:ell380}c.
As it is shown in Fig.~\ref{fig:ell380}d, a reduction of $t_Z$ is observed as well, when $\ell$ increases: $t_Z$ is 7 years for $\ell=190\, \textrm{km}/\textrm{week}$, whereas it is around 1.2 years for $\ell=380\, \textrm{km}/\textrm{week}$.
The equilibrium expenses at $t_X$ have been slightly reduced to $X\approx 5.06\times10^5\, \$\textrm{MXN}$, while the equilibrium loss at $t_Z$ has significantly decreased to $Z\approx 1.34\times10^5\, \$\textrm{MXN}$.

The effects of the driving distance $\ell$ and the congestion coefficient $\alpha$ over the equal expenses time $t_X$ and losses time $t_Z$, and the corresponding equilibrium expenses $X$ and losses $Z$, have been studied.
The results, considering the same average speed $U=25\, \textrm{km}/\textrm{h}$, are depicted in Fig.~\ref{fig:elldependence}.
Increasing $\ell$ leads to a reduction of the equal expenses time $t_X$ and the corresponding equilibrium value $X$, stepper changes for small values of $\ell$ and more gentle for large values of $\ell$, as shown in Figs.~\ref{fig:elldependence}a and ~\ref{fig:elldependence}b.
As the driving distance in a week $\ell$ grows, the economic benefits of an electric vehicle stand out.
Three different values of $\alpha$ have also been explored, indicating $\alpha=1$ for light, $\alpha=1.75$ for average and $\alpha=2.5$ for heavy traffic.
As $\alpha$ grows, the curves for $t_X$ and $X$, both are vertically shifted towards lower times and expenses, meaning that the advantages of an electric vehicle are strengthened as traffic conditions increase.

Decreasing magnitudes are also observed for the equal losses time $t_Z$ and the corresponding equilibrium value $Z$, either for increasing values of $\ell$ or larger values of $\alpha$, as depicted in Figs.~\ref{fig:elldependence}c and ~\ref{fig:elldependence}d.
Huge values of $t_Z>10\, \textrm{years}$ and $Z>6\times10^5\, \$\textrm{MXN}$ are discerned for $\ell<200\, \textrm{km}/\textrm{week}$ and $\alpha=1$, which are far beyond the expected lifespan of a vehicle and attaining the price of a second vehicle.
Under this conditions, the acquisition of an electric vehicle is strongly not advised.
In contrast, for $\alpha=2.5$, small values of $t_Z<5\, \textrm{years}$ and $Z<3\times10^5\, \$\textrm{MXN}$ are found for $\ell>150\, \textrm{km}/\textrm{week}$, which are reasonable times and affordable quantities that make an electric vehicle a very interesting choice.
For $\alpha=1.75$, the same advantageous conditions of an electric vehicle are attained for $\ell>200\, \textrm{km}/\textrm{week}$.

\section{Conclusions}

When the acquisition of a new vehicle is required, deciding whether to buy a gasoline or an electric one is a complex task.
The choice depends on many factors: 1) the personal economic conditions to provide an initial investment, 2) the personal driving characteristics and habits, 3) the needs of the user (driving distance), 4) the local speed limits, 5) the in-place traffic conditions, and 6) the economic behavior of the country, among other situations.
Additionally, the time that the user pretends to keep the vehicle may be an important variable to evaluate the options.

Our analysis has been performed considering the price of a new vehicle (initial investment), the operation cost (gasoline refueling or battery recharging), the maintenance and services costs, and the local taxes, all of them taking into account the evolution in time of prices (some of them due to inflation).
If the future owner will use the vehicle for a driving distance per week that is relatively short (below $200\, \textrm{km}/\textrm{week}$) buying a gasoline vehicle seems like a good option, when traffic conditions are light.
However, for heavy traffic conditions, as it is usually the case observed in large cities, an electric vehicle is the option that brings economical benefits in the long term (after 5 years of use).
Another important consideration is whether the future owner would like to keep the vehicle through its entire lifespan or would like to sell it in the short time.
If the choice is to keep it, the economical advantage of an electric vehicle will arrive sooner, for high vehicle usage, or later, for low vehicle usage.
Making the choice of selling it, 
low usage of an electric vehicle brings a capital loss, if the choice is to sell it quickly (prior to 5--6 years of use).
In contrast, a high vehicle usage allows to sell it, even from the second year of use, and yet save money.
Additionally, the technological advances that will be surely achieved in the future, due to the research and development in energy storage systems, will reduce the initial cost of electric vehicles, shortening the time for which expenses of both vehicles become similar.

For situations in which $\ell>383.6 \, \textrm{km}/\textrm{week}$, exceeding what we consider as a large weekly driving distance in this work, the frequency of services and maintenance must be adjusted.
A new frequency for these expenses should be calculated, based on the determination of the time at which a 10000 km mileage is covered. 
Additionally, the application of inflation for these expenses needs to be adjusted.
Cab and delivery services cover distances exceeding $383.6 \, \textrm{km}/\textrm{week}$, thus falling into this category.
Despite the increased maintenance frequency, electric vehicles should presents greater economic benefits, in both energy consumption and maintenance costs.
The initial investment should be recovered more quickly, and the economic benefits should be attained in shorter times.
A future work, including the corresponding modifications to our model, will be presented as an extension of this study, addressing the case of large weekly driving distances.

Besides the economical considerations, we have to take into account the fact that in some countries there are pollution policies.
This policies may cause that gasoline vehicles can not be used everyday, when pollution indices are high.
Under these conditions, an electric vehicle seems to be an interesting option.
Nevertheless, a more extensive analysis must include the social value of emissions and pollutants.
In the present work, the pollutant emissions from electricity generation are not taken into account, which is a factor that must be considered to observe the full picture. 
Likewise, an economic value for human time must also be accounted for in a complete study.
For instance, the gasoline refueling may be time consuming, whereas battery recharging can be performed when other activities take place, such as sleeping, visiting a mall, etc..
Despite all the first-sight advantages of electric vehicles, there are still conditions under which gasoline vehicles may be the optimal selection.
Our main conclusion is that a complete analysis, considering the economical trends of all the variables, must be done before making a choice for buying a new vehicle.

\section{Data availability statement}

The data that support the findings of this study are openly available.
Data can be downloaded directly from the websites listed in the references~\cite{GasolinaMX,ElectricidadMX,DepreciacionMX,InflationMX,NissanMX,JacMX,UMAMX,TenenciaMX,EVMX}.

\section{References}

\bibliographystyle{unsrt}
\bibliography{Vehicles_v2}

\end{document}